\documentstyle[multicol,prb,aps,epsfig]{revtex}

\begin{document}

\twocolumn[\hsize\textwidth\columnwidth\hsize\csname @twocolumnfalse\endcsname

\title{Persistent currents in mesoscopic rings: A numerical and
      renormalization group study}

\draft

\author{V.\ Meden}
\address{Institut f\"ur Theoretische Physik, Universit\"at G\"ottingen, 
Bunsenstr.\ 9, 
D-37073 G\"ottingen, Germany}
\author {U.\ Schollw\"ock}
\address{Max-Planck-Institut f\"ur Festk\"orperforschung,
  Heisenbergstr.\ 1, D-70569 Stuttgart, 
Germany} 

\date{September 25, 2002}
\maketitle

\begin{abstract}
The persistent current in a lattice model of a one-dimensional
interacting electron system is systematically
studied using a complex version of the density matrix renormalization
group algorithm and 
the functional renormalization group method. We mainly focus on the 
situation where a single 
impurity is included in the ring penetrated by a magnetic flux. Due to the interplay of the
electron-electron interaction and the impurity the persistent current in a system of $N$ 
lattice sites vanishes faster then $1/N$. Only for very large systems
and large impurities 
our results are consistent with the bosonization prediction obtained for an effective 
field theory. The results from the density matrix renormalization
group and the functional renormalization group agree
well for interactions as large as the band width, even though as an approximation 
in the latter method the flow of the two-particle vertex is neglected. This confirms that 
the functional renormalization group method is a very powerful tool to
investigate correlated electron systems. The method will become very
useful for the
theoretical description of the electronic properties of small conducting ring molecules.
 \end{abstract}


\vskip 2pc]
\vskip 0.1 truein
\narrowtext
\section{Introduction}
\label{secintro}

The experimental observation of persistent currents in mesoscopic metallic and 
semiconducting rings pierced by a magnetic flux\cite{Levy,Chandra,Mailly,Reulet,Jariwala,Rabaud} 
has led to many theoretical 
investigations focusing on the interplay of electron-electron interaction and 
disorder in such systems.\cite{pctheory} This interplay is considered
to be one of the possible reasons 
for the large current observed in the experiments.\cite{pctheory}
Despite these studies a 
quantitative theoretical understanding of the observed amplitude of the 
currents for the three-dimensional rings is still missing.
To gain theoretical insight the simplified situation of one-dimensional rings
with interaction and disorder was studied using exact diagonalization
for systems of very few lattice sites (up to 16)\cite{Abraham,didier,Kato} and the
self-consistent Hartree-Fock approximation.\cite{Kato,Cohen} 
Here we consider the further simplified  
problem of the persistent current in a one-dimensional ring of interacting electrons 
in the presence of a single impurity and penetrated by a magnetic flux. 
Within an effective continuum field theory it
has theoretically been investigated using
bosonization\cite{gogo} and conformal field theory.\cite{malte,Jaimungal} 
At first glance this problem seems to be of purely academic interest, but the
fast progress in the design and manipulation of conducting 
ring molecules gives a perspective that such systems might be accessible to 
experiments in the very near future. 

For a continuum model
of non-interacting one-dimensional electrons the leading 
behavior of the persistent current $I(\phi)$ in the system size $L$ can be 
calculated in the presence of an arbitrary potential scatterer.\cite{gogo} 
It is a periodic function of the flux, vanishes as $1/L$, 
and its shape and size are determined by the 
absolute value of the transmission amplitude $|T(k_F)|$ of the potential at the Fermi wave 
vector $k_F$. For a 
vanishing impurity, i.e.\ $|T(k_F)| \to 1$, $I(\phi)$ has a saw tooth like shape, which gets 
rounded off if $|T(k_F)|$ is decreased. In the limit of small $|T(k_F)|$, $I(\phi)$ is 
proportional to $|T(k_F)| \sin{\phi}$. For the tight-binding lattice model supplemented 
by a single weak hopping matrix element the persistent current at half-filling can also be calculated 
analytically\cite{malte} and the same characteristics can be 
found (see Sect.\ \ref{secmodel}). 

Compared to the Fermi liquid behavior of higher dimensional systems
a large class of models of homogeneous one-dimensional interacting
electrons has a significantly different low-energy physics. These models belong to the 
Luttinger liquid universality class.\cite{Haldane1} The low-energy excitations of Luttinger 
liquids are not given by fermionic quasi-particles, but are of collective, bosonic nature.
This leads e.g.\ to a typical power-law  decay of correlation functions. 
The low-energy physics of 
Luttinger liquids is characterized by a set of interaction and filling factor dependent
parameters.\cite{Haldane1} For the case of spinless fermions on which we focus any pair of 
the four parameters $v_J$ (velocity of current excitations),
$v_N$ (velocity relevant if particles are added), $v_{c}$ (velocity of 
charge excitations at constant number of particles), and the Luttinger
liquid parameter $K$ can be used. Within bosonization
and for  
the impurity free case the persistent current in a Luttinger liquid is
periodic in $\phi$ and of saw tooth like shape
with slope $v_J/(\pi L)$.\cite{Haldane1,Schick,Loss} 
In Sect.\ \ref{secdmrg} we will compare our numerical results to this prediction.

The low-energy physics of Luttinger liquids is  strongly affected
by the presence of a single 
impurity.\cite{LutherPeschel,Mattis,ApelRice,KaneFisher,EggertAffleck,MatveevGlazman}
The problem is usually mapped onto an effective continuum field theory 
using bosonization, where terms which are expected to be irrelevant 
in the low-energy limit are 
neglected.\cite{LutherPeschel,Mattis,ApelRice,KaneFisher,EggertAffleck}
Within this field theory the leading $L$ dependence of the persistent
current was obtained by an additional self-consistent approximation 
using the analogy to the problem of quantum coherence in a
dissipative environment.\cite{gogo} This approach gives a current
which for large $L$ vanishes as $L^{-\alpha_B-1}$, with
$\alpha_{B}=1/K-1$, and independent of
the bare impurity strength is of purely sinusoidal shape.\cite{gogo} 
For repulsive interaction one has $K<1$ and thus $\alpha_B>0$.
$\alpha_B$ is also the 
exponent of the power-law suppression (as a function of energy) of the local spectral weight 
close to an open boundary and the chemical potential.\cite{KaneFisher,VM2} This explains the  
index $B$ which stands for boundary. 
Many of the bosonization results for observables which are dominated
by the interplay of a single impurity and the electron-electron 
interaction can be understood in terms of a single particle
picture, which for Luttinger liquids has of course 
to be used with caution: For large $L$ 
the effective transmission amplitude 
near the Fermi points is suppressed with respect to the non-interacting 
transmission by a factor of $L^{-\alpha_B}$. Combining this with the
behavior of the persistent current in the 
non-interacting case the above result, obtained without the use of 
the single particle language, can be derived.


For several reasons it is desirable to directly show the above behavior of the
persistent current in microscopic lattice models avoiding bosonization. 
Mapping such models on the field theory involves 
approximations. Their validity can be questioned and they lead to
a loss of information; the scales of the microscopic model and 
corrections to the expected power-law scaling hidden in the irrelevant
terms are lost. Furthermore even  within the field theory an
additional 
approximation is necessary to determine the leading (in the
system size) behavior of the persistent current. 
A knowledge of the detailed shape of $I(\phi)$ beyond the leading
behavior and for microscopic models is of 
special importance if one wants to compare theoretical results to experiments.  
Several attempts have been made in this direction using the model 
of spinless fermions with nearest neighbor hopping and interaction.
Here we also focus on this model. The persistent current
can be calculated by taking the derivative of the groundstate energy $E_0(\phi)$ with respect 
to the flux $\phi$ penetrating the ring. Instead of calculating the full functional form of
$I(\phi)$ the so called phase 
sensitivity $\Delta E_0$, which is the difference of the groundstate energy at flux $0$ and 
$\pi$ has numerically been determined using the density matrix renormalization
group (DMRG) method.\cite{Peterdok,Peters1} The phase sensitivity can
be considered  a crude measure for the persistent current, 
but of course does not contain information on the detailed shape of the current as a 
function of $\phi$. Very recently $\Delta E_0$ has again been studied
numerically using DMRG and the quantum Monte Carlo method.\cite{Byrnes}
$\Delta E_0$ instead of $I(\phi)$ (which implies calculating $E_0(\phi)$ 
for several $\phi$ and numerically taking the derivative) is calculated because this way 
the hamiltonian matrix remains real and the numerical effort is reduced 
considerably.\cite{footnote1} 
Here we use complex DMRG to calculate $I(\phi)$ with very high precision and for systems of
up to $N=128$ lattice sites.\cite{footnote2} Additionally the
functional renormalization group (RG) method introduced 
recently into the theory of strongly correlated electrons is
used.\cite{Wetterich,Morris} 
It has been applied to  two-dimensional correlated electron
systems\cite{2dsystems} and 
one-dimensional Luttinger liquids.\cite{VMa,VMb,Tom} 
We have used this method before to study a local observable\cite{VMa,VMb} - the local
spectral weight close an impurity - which is also dominated by the
interplay of electron-electron interaction and impurity. In contrast
the persistent current is a property of the entire system.  
Within the RG approach the
flow equations are 
closed by neglecting the flow 
of the two-particle vertex. Nonetheless DMRG and RG agree quantitatively for interactions 
of the order of the band width. This confirms that 
the functional RG is a very powerful tool to investigate 
strongly correlated electrons. For the
single impurity case we indeed find 
that the persistent current vanishes faster then $N^{-1}$. To analyze $N I(\phi)$ in more 
detail we expand the current in a Fourier 
series, demonstrate that for large impurities and very large system sizes the behavior 
of the first Fourier coefficient is consistent with a power-law decay with exponent 
$-\alpha_B$, and that the higher order coefficients decay even faster. 
Thus in the $N \to \infty$ limit $ I(\phi)$ is proportional to $\sin{\phi}$ with 
an amplitude which vanishes as $N^{-\alpha_B-1}$. However, this 
universal bosonization prediction only holds for very long chains respectively
very large impurities. For smaller systems and impurities
the asymptotic limit is not reached and the current displays a more
complex behavior. It can quantitavely be described
using the functional RG method. This makes the method an ideal tool to
investigate the electronic properties of small, one-dimensional molecular rings 
which might be experimentally accessible very soon.

\section{The Model}
\label{secmodel}

The Hamiltonian for the impurity free ring  penetrated by a magnetic
flux $\phi$ (measured in units of the flux quantum $\phi_0=h c/e$) 
with nearest neighbor interaction $U$ is given by 
\begin{eqnarray}
H & = & - \sum_{j=1}^{N} \left( c_j^{\dag} c_{j+1}^{} e^{i \phi/ N}+
  c_{j+1}^{\dag} c_j^{} e^{-i \phi/ N} \right) \nonumber \\ && + U \sum_{j=1}^{N} n_j n_{j+1} \; ,
\label{spinlessfermdef}
\end{eqnarray}
in standard second-quantized notation. The hopping matrix element and the lattice constant 
are set to one  and periodic boundary conditions are used. 
For $U=0$ the groundstate energy of the model can easily be calculated.
At temperature $T=0$ and for fixed $N$ the persistent current follows from  
$E_0(\phi)$ by taking the derivative with respect to $\phi$ 
\begin{eqnarray}
I(\phi) = - \frac{d E_0(\phi)}{d \phi}
\label{currentdef}
\end{eqnarray} 
as can be shown using the Hellman-Feynman theorem. To leading order in $1/N$ this gives 
\begin{eqnarray*}
I(\phi) = - \frac{v_F}{\pi N} \times \left\{
            \begin{array}{cl}
            \phi  & \mbox{for} \, N_F \, \mbox{odd and} \, -\pi \leq \phi < \pi \\
           \phi-\pi  & \mbox{for} \, N_F \, \mbox{even and} \, 0 \leq
           \phi < 2 \pi \; ,
\end{array} \right. 
\end{eqnarray*} 
where $v_F$ denotes the Fermi velocity and $N_F$ is the number of
particles.\cite{Cheung} Both functions have 
to be continued periodically. 
Eq.\ (\ref{currentdef})  also holds for non-vanishing interaction
and if impurity terms are added.
The above even-odd effect can also be
observed if interaction and impurities are 
included. We from now on only considere even $N_F$. 

At $\phi=0$ 
the Luttinger liquid parameter $K $ and the velocity of current excitations $v_J$ 
of the model Eq.\ (\ref{spinlessfermdef}) can be determined
using  the Bethe ansatz.\cite{Haldane2} For half-filling the resulting integral equations 
have been solved analytically with the results
\begin{eqnarray} 
K =\left[ \frac{2}{\pi} \arccos{ \left( - U/2 \right)}
\right]^{-1} 
\label{Krhohalf}
\end{eqnarray}
and ($v_J=v_c K$)
\begin{eqnarray} 
v_J= \pi \frac{\sqrt{1-(U/2)^2}}{\pi - \arccos{\left( -
      U/2 \right)}} \left[ \frac{2}{\pi} \arccos{ \left( - U/2 \right)}
\right]^{-1}  \; .
\label{vJhalf}
\end{eqnarray}
For this filling the model is a Luttinger liquid for $-2 < U \leq
2$. At $|U|=2$ the model shows phase transitions to a charge density wave groundstate
($U=2$) and a phase separated state ($U=-2$).  
To leading order in $1/N$ the bosonization prediction for the persistent 
current in a homogeneous Luttinger liquid is a saw tooth 
like curve with slope $-v_J/(\pi N)$, i.e.\ compared to the non-interacting case $v_F$ is replaced
by the velocity determining the current excitations.
To compare our data with this result also away from half-filling we numerically 
solved the Bethe ansatz integral equations following Refs.\
\onlinecite{Haldane2} and \onlinecite{Qin} and 
determined $v_J$ (see Sect.\ \ref{secdmrg}). In our study we will always stay in the Luttinger
liquid phase.

To $H$ we 
add a hopping impurity 
\begin{eqnarray}
H_h = (1-\rho) \left( c_{N}^{\dag}
c_{1}^{} e^{i \phi/ N} + c_{1}^{\dag}
c_{N}^{} e^{-i \phi/ N} \right) \; .
\label{hopimpdef}
\end{eqnarray}
with $\rho$ between $0$ (no hopping between sites $N$ and $1$) and $1$ (no impurity). 
In the non-interacting limit, at half-filling, and at wavevector $k_F$
the transmission coefficient of the 
hopping impurity (for $N \to \infty$) is given by\cite{VM3}
\begin{eqnarray}
|T(k_F)|^2= \frac{4\rho^2}{(1+\rho^2)^2} \; ,
\label{Tkfdef}
\end{eqnarray} 
which provides us with a measure for the strength of the impurity.
For $U=0$, $N_F=N/2$, and to leading order in
$1/N$ the persistent 
current of the hamiltonian $H+H_h$ has been calculated using
conformal field theory.\cite{malte}  To obtain $I(\phi)$ it is not necessary to use
this field theoretical approach since Eq.\ (7) of Ref.\
\onlinecite{malte} which determines the allowed wave vectors can be
solved directly. Written in terms of $|T(k_F)|$ the current is given by 
\begin{eqnarray}
I(\phi) = \frac{v_F}{\pi N}  \frac{\arccos{\left( |T(k_F)|
      \cos{\left[\phi -\pi \right]} \right) }}{\sqrt{1- |T(k_F)|^2
    \cos^2{\phi }}} |T(k_F)| \sin{\phi} \; , 
\label{U0current}
\end{eqnarray}
which is exactly the expression obtained in Ref.\ \onlinecite{gogo} 
for the continuum model.  
Later we will be interested in the small $|T(k_F)|$ limit. 
Therefore we expand Eq.\ (\ref{U0current}) up to third order
in $|T(k_F)|$
\begin{eqnarray}
 N I(\phi)  & = & \frac{v_F}{2} |T(k_F)| \left( 1+ |T(k_F)|^2 /
  8\right) \sin{(\phi)} \nonumber \\ && + \frac{v_F}{2\pi}
|T(k_F)|^2 \sin{(2 \phi)} \nonumber \\ && + \frac{v_F}{16} |T(k_F)|^3 \sin{(3
  \phi)} 
  + {\mathcal O} \left( |T(k_F)|^4 \right) \; ,
\label{U0currentexp}
\end{eqnarray} 
which at the same time gives an expansion in a Fourier series. The
expansion explicitly shows that in the limit $|T(k_F)| \to 0$ of a strong impurity  
the current becomes more and more of sinusoidal shape.

A simple approximation which allows to study the effect of the interaction
and impurity on the persistent current simultaneously 
is the Hartree-Fock approximation. For the bulk properties of
homogeneous one-dimensional correlated electron systems this
approximation does not capture any of the Luttinger liquid features   
and is thus of very limited usefulness. Furthermore, when applying the 
self-consistent Hartree-Fock approximation to a model with a single 
impurity, the self-consistent iterative solution of the Hartree-Fock
equations will drive the system into a charge
density wave groundstate with a finite single-particle gap which is
qualitatively incorrect since a single 
impurity cannot change bulk properties of the system.   
Nevertheless self-consistent 
Hartree-Fock has been 
used to determine the persistent current in one-dimensional,
interacting, and disordered 
rings\cite{Kato,Cohen} and also for the single impurity case.\cite{Cohen}
In contrast if one is interested in the local properties close to a
boundary or impurity 
non-self-consistent Hartree-Fock provides useful informations.\cite{VM2,VMa,VMb}
In Sect.\ \ref{secresults} we also present results for the
persistent current calculated within the non-self-consistent Hartree-Fock
approximation. They were  obtained by numerically determining the
groundstate of $H+H_h$ for $U=0$ and fixed $\phi$ and $N$. From this 
the expectation values $\left< n_j \right>_{0} $ and 
$\left< c_{j+1}^{\dag} c_j \right>_{0}$, which determine the
mean-field hamiltonian $H_{\rm MF}$, can be
calculated. $H_{\rm MF}$ can then be diagonalized numerically. There are two
possibilities to determine the groundstate energy within this
approximation. One can either take the Slater determinant expectation value
of $H+H_h$ using the groundstate of $H_{\rm MF}$ or determine the
energy via the one-particle propagator\cite{AGD} by the formula
\begin{eqnarray}
\left< H+H_h \right>_{\rm MF} & = & \frac{1}{2} \Biggl[ - \lim_{\tau' \to
    \tau} \sum_{j=1}^{N}  \left< c_{j}^{\dag}(\tau') \frac{d}{d \tau}
  c_j(\tau) \right>_{\rm MF}   \nonumber \\*
 && -  \sum_{j=1}^{N} \left(e^{ i \phi/N } \left<
  c_{j}^{\dag} c_{j+1} \right>_{\rm MF}  + \, \mbox{c.c.} \, \right)  \nonumber
\\*  &&  + (1-\rho) \left(e^{ i \phi/N } \left<
  c_{N}^{\dag} c_{1} \right>_{\rm MF} + \, \mbox{c.c.} \, \right)
  \Biggr] 
\label{energyfromprop}
\end{eqnarray} 
where the expectation values are taken using the groundstate of $H_{\rm
  MF}$. $ c_{j}^{(\dag)}(\tau)$ denotes the annihilation (creation)
operator in the imaginary time Heisenberg representation. Transforming
the first term into the Matsubara frequency representation 
it can be written as 
$\sum_{l=1}^{N_F} \varepsilon_{l}^{\rm MF}$, with the eigenenergies
$\varepsilon_{l}^{\rm MF}$ of $H_{\rm MF}$.
Only if the self-consistent Hartree-Fock approximation is
used, which for the reason given above we do not consider in the
present context,
both possibilities give the same result. Formally both approximations to the energy 
are correct to leading order in $U$, but it turned out that the latter method
gives better results compared to the high-precision DMRG
data. For the results presented in Sec.\ \ref{secresults} we thus used
Eq.\ (\ref{energyfromprop}).
There it will be shown that the currents calculated using the Hartree-Fock 
approximation are qualitatively wrong, since in this method 
correlation effects are neglected which in the present context are of great importance. 

\section{Complex DMRG}
\label{secdmrg}

Using the DMRG algorithm the groundstate energy of an interacting
one-dimensional many fermion system can be calculated to high
precision.\cite{DMRGbasics} To determine the persistent current using 
Eq.\ (\ref{currentdef}) for the hamiltonian $H+H_h$ [Eqs.\ (\ref{spinlessfermdef})
and (\ref{hopimpdef})] the DMRG procedure has to be generalized to 
complex hamiltonian matrices. Calculation time scales up by about a factor of 4, memory usage by a factor of 2. This limits the performance of the method,
which is however numerically very stable as hermiticity is conserved. We have
kept up to 400 states, ensuring that energies and derived currents are essentially exact.

\begin{figure}[hbt]
\begin{center}
\vspace{0.5cm}
\leavevmode
\epsfxsize7.7cm
\epsffile{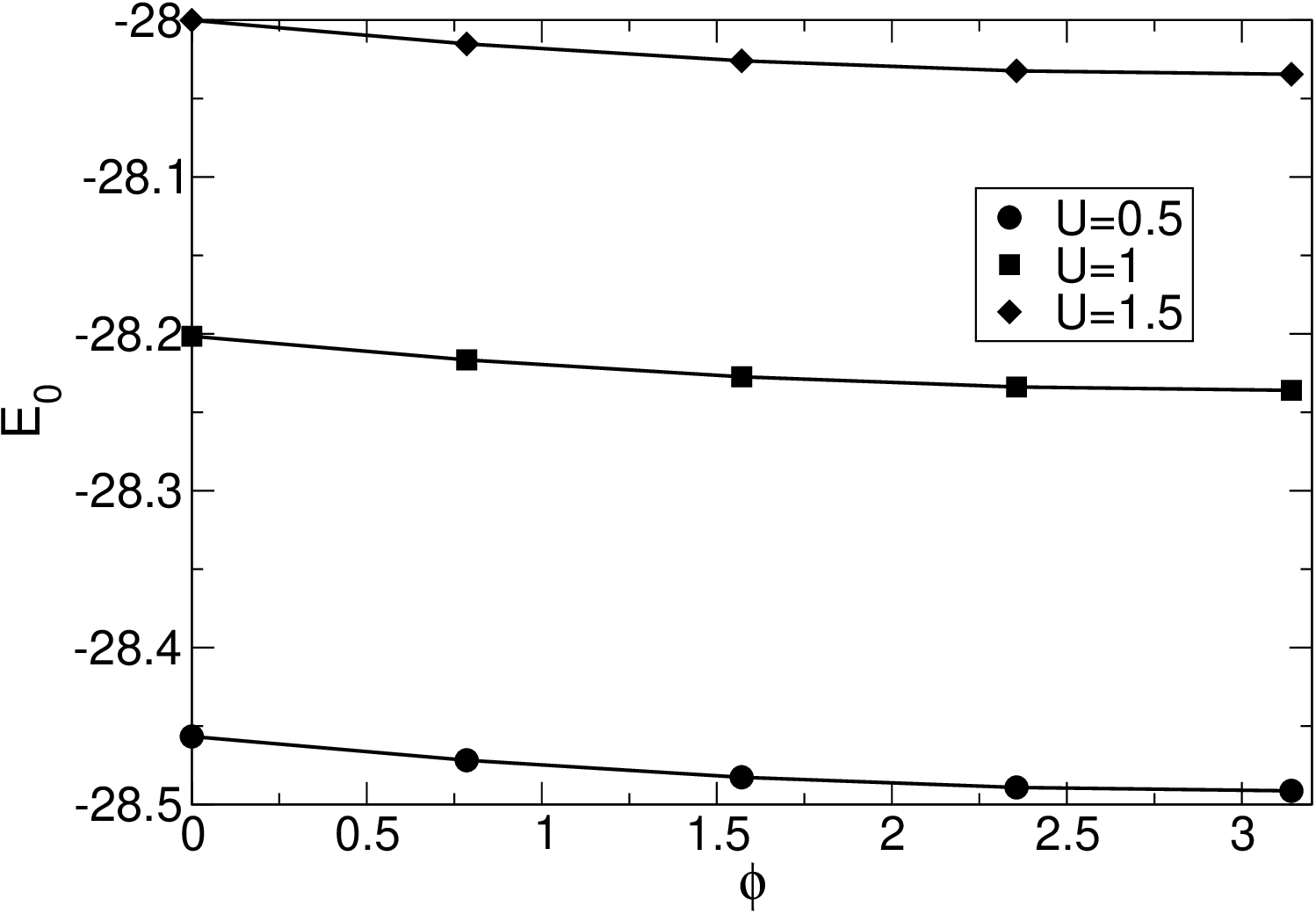}
\caption{Groundstate energy as a function of the flux $\phi$ for
  quarter-filling, $N=64$, and different $U$. The symbols are DMRG
  data and the lines are quadratic fits (see the text).}
\label{fig1}
\end{center}
\vspace{-0.0cm}
\end{figure}

\begin{figure}[hbt]
\begin{center}
\vspace{0.0cm}
\leavevmode
\epsfxsize7.7cm
\epsffile{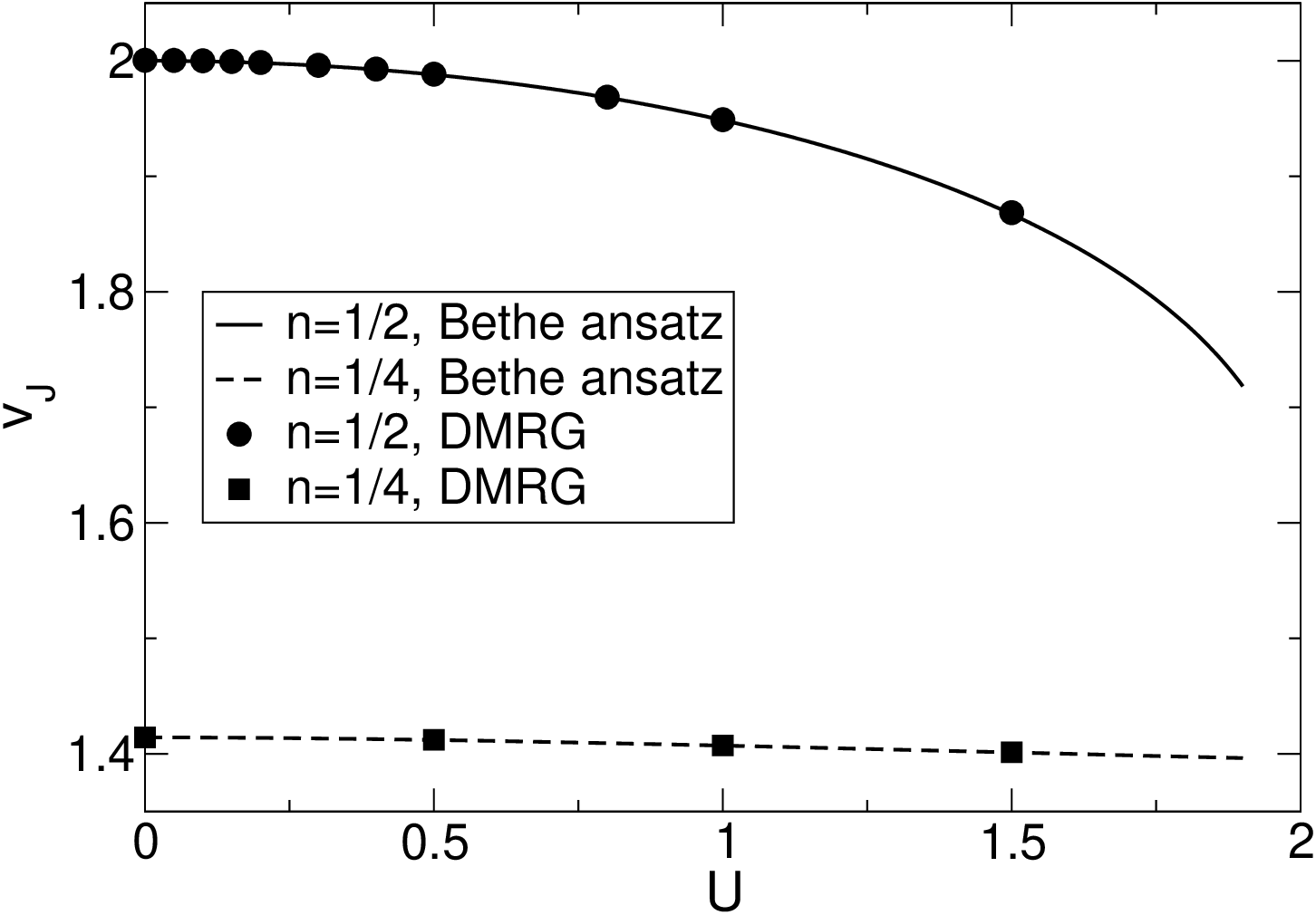}
\caption{Current velocity $v_J(n,U)$ as obtained from the Bethe ansatz and
  from the DMRG.}
\label{fig2}
\end{center}
\vspace{-0.0cm}
\end{figure}

As a test of our program we first studied the impurity free case given
by the hamiltonian Eq.\ (\ref{spinlessfermdef}). We calculated the groundstate
energy as a function of $\phi$ for $0 \leq \phi \leq \pi$. Results for
quarter-filling $n=1/4$, 
$N=64$ and $U=0.5,1,1.5$ are shown in Fig.\ \ref{fig1}. Bosonization
predicts that to leading order in $1/N$ the current
is given by
\begin{eqnarray} 
I(\phi) = -\frac{v_J}{\pi N} (\phi-\pi)
\label{boscurrent}
\end{eqnarray}
and thus the groundstate energy by 
\begin{eqnarray} 
E_0(\phi)=\mbox{const.} \, + \frac{v_J}{2 \pi N} (\phi-\pi)^2 \; .  
\label{bosenergy}
\end{eqnarray}
As shown in Fig.\ \ref{fig1} the DMRG data nicely lie on 
curves of this form  with
the constant and the current velocity as fit parameters. In Fig.\
\ref{fig2} the $v_J^{\rm DMRG}(n,U)$ extracted from such fits are
compared to the exact $v_J(n,U)$ obtained from the Bethe ansatz. Data for 
$n=1/4$ (with $N=64$) and $n=1/2$ (with $N=60$) and different $U$ are shown. 
For both fillings the Bethe ansatz and DMRG results are
indistinguishable on the scale of the plot. The relative error of the
DMRG data is of the order of $10^{-4}$. This confirms the bosonization
prediction, shows that the complex
DMRG can be applied successfully, and that the current velocity can to very
high precision be extracted from finite size data as small as $60$
lattice sites without using finite size scaling. In Ref.\ \onlinecite{Peterdok}
data for the phase sensitivity of the translational invariant model at
half-filling were obtained using DMRG. Our results for  
$v_J^{\rm DMRG}(n=1/2,U)$ are consistent with the ones presented there.
As seen in Fig.\ \ref{fig2} the dependence of $v_J$ on $U$ gets weaker
if the filling is getting smaller. This happens  because for smaller
fillings the lattice model is closer to the electron gas model with
quadratic dispersion. The latter model is Galilean-invariant which
leads to $v_J=v_F$ independent of the interaction. 

\section{RG method}
\label{secrg}

Besides DMRG we will also use the functional RG method in the version 
using one-particle irreducible vertex 
functions.\cite{Wetterich,Morris} In collaboration with W.\ Metzner
and K.\ Sch\"onhammer we have successfully applied this method in the past 
to determine the local spectral weight of a Luttinger liquid close to
an open boundary and an impurity.\cite{VMa,VMb} In the method one
introduces a cut-off parameter $\Lambda$ in the non-interacting propagator $G^0$
cutting out degrees of freedom on energy scales less than $\Lambda$
and derives an exact hierarchy of coupled differential flow equations
for the one-particle irreducible vertex functions by differentiating
with respect to $\Lambda$, where $\Lambda$ flows from $\infty$ to $0$.

For $\phi=0$ the flow equation for the selfenergy\cite{footnote3} of
the hamiltonian $H+H_h$ has been given 
in Ref.\ \onlinecite{VMb}. As in this reference we here also neglect 
the flow of the two-particle vertex which closes the set of differential
equations; this leads to a energy independent selfenergy.
Within the approximation the results obtained are at least
correct to leading order in $U$, but our work presented in Refs.\
\onlinecite{VMa} and \onlinecite{VMb} shows that the
Luttinger liquid scaling of the impurity (i.e.\ the transmission) is
included,  which makes the RG a promising method also in
the present context. 

If a frequency cutoff 
\begin{equation}
 G^{0,\Lambda}(i\omega) = \Theta(|\omega| - \Lambda) 
G^{0}(i\omega) 
\label{g0deff}
\end{equation}
is used, the set of flow equations for the
selfenergy\cite{VMb} in the Wannier basis is given by
\begin{eqnarray}
\frac{d}{d \Lambda}\Sigma^{\Lambda}_{j,j} & = & - \frac{U}{2\pi} 
  \sum_{s=\pm 1} \sum_{\omega = \pm \Lambda}
  G^{\Lambda}_{j+s,j+s} (i \omega) 
  \label{diffsystem1} \\
\frac{d}{d \Lambda} \Sigma^{\Lambda}_{j,j\pm 1} & = & \frac{U}{2\pi} 
  \sum_{\omega = \pm \Lambda}
  G^{\Lambda}_{j,j \pm 1} (i \omega)  \; ,
\label{diffsystem}
\end{eqnarray}  
with 
\begin{eqnarray}
G^{\Lambda}(i\omega) = \left\{ \left[ G^0( i \omega) \right]^{-1} -
  \Sigma^{\Lambda} \right\}^{-1} \; .
\label{fullgdef}
\end{eqnarray}
In Eqs.\ (\ref{diffsystem1}) and (\ref{diffsystem}) the flow from
$\Lambda=\infty$ down to a scale $\Lambda_0$ much larger then the 
band width has already been included. The flow is continued from 
$\Lambda_0$ downwards with the initial conditions 
\begin{eqnarray}
\Sigma^{\Lambda_0}_{j,j} & = & U \;\;\; ,  \;\;\; 1\leq j \leq N
\nonumber \\ 
\Sigma^{\Lambda_0}_{j,j \pm 1} & = & 0 \;\;\; ,  \;\;\; 1 < j <  N
\nonumber \\ 
\Sigma^{\Lambda_0}_{1,N} & = & (1-\rho) \; e^{-i \phi/N} = 
\left(\Sigma^{\Lambda_0}_{N,1}\right)^{\ast} \; .
\label{initialcond}
\end{eqnarray}
In contrast to the situation studied in Refs.\ \onlinecite{VMa} and
\onlinecite{VMb} the matrix elements $\Sigma^{\Lambda}_{j,j\pm 1}$ are
complex numbers
which increases the size of the set of differential
equations. Furthermore we here have to use periodic boundary
conditions so that $G^{\Lambda}(i\omega)$ is not tridiagonal. 

The set of Eqs.\ (\ref{diffsystem1}) and (\ref{diffsystem}) is complemented by a
differential equation for the ``zero-particle'' vertex $\gamma_0$
(with the Boltzmann constant $k_B$ set to one)
\begin{eqnarray}
\lim_{T \to 0}  T \frac{d}{d \Lambda} \gamma_0^{\Lambda} = &&
\nonumber \\
\frac{1}{2
  \pi} \, \mbox{Tr} && \left[ \sum_{\omega = \pm \Lambda} \ln{\left\{
      1- G^0(i \omega) \Sigma^{\Lambda}(i \omega) \right\} }  \right] \; ,
\label{gamma0flow}
\end{eqnarray}  
where $\mbox{Tr}$ denotes the trace over the indices of the Wannier
basis states. The initial condition is $\gamma_0^{\Lambda=\infty} =
\gamma_0^{\Lambda_0}=0$. 
Eq.\ (\ref{gamma0flow}) also holds if the flow of the
two-particle vertex and all higher vertices are taken into account. 
The flow of $\gamma_0^{\Lambda}$ only couples to
the selfenergy. It does not feed back to any higher order vertex function.   
$\gamma_0=\gamma_0^{\Lambda=0}$ is related to the grand canonical potential $\Omega$ by
\begin{eqnarray}
\Omega = T \gamma_0 + \Omega_0 \; ,
\label{Omegadef}
\end{eqnarray} 
where $\Omega_0$ is the grand canonical potential at $U=0$. 
From $\Omega$ the ground state energy can be obtained in the $T \to 0$
limit
\begin{eqnarray}
E_0(\phi) - \mu \left< \hat N \right> & = & \Omega(T=0) \nonumber \\ 
& = &  \lim_{T \to 0} T \gamma_0 + E_0^0(\phi) - \mu_0 \left< \hat N \right>_0 \; .
\label{energyfromOm}
\end{eqnarray} 
$\hat N$ is the particle number operator and $\mu$ the chemical
potential. The quantities with an additional index $0$ are taken at 
$U=0$. Eqs.\ (\ref{Omegadef}) and (\ref{energyfromOm}) make explicit that the 
functional RG is a
grand canonical method. In general to obtain the persistent current
within a grand canonical ensemble corrections to Eq.\ (\ref{currentdef}) have to
be taken into account since the chemical potential can dependent on
the flux.\cite{Schmid,Altshuler} For half-filling and a single hopping
impurity $\mu$ is always fixed at $U$ (and $\mu_0$ at $0$) 
and thus we do not have to worry about this issue if we use the RG only for $n=1/2$. 

The set of differential equations (\ref{diffsystem1}),
(\ref{diffsystem}), and (\ref{gamma0flow}) can be solved
numerically for a fixed $N$ and $\phi$. Additionally $E_0^0(\phi) $
can be calculated by
numerically diagonalizing the one-particle problem at $U=0$. Thus
$E_0(\phi) $ can be determined. There are two factors
which limit the system sizes that can be treated numerically. On the right hand
side of the flow equations a $N \times N$ matrix has either to be
inverted or to be diagonalized. This limits $N$ to a few thousand
lattice sites. The more restrictive limit comes from the fact that
within the ground state energy which is proportional to $N$ 
one is interested in the $\phi$ dependence which vanishes faster 
than $1/N$ (see above and the next section). Thus the differential 
equations have to be solved with extremely high precision.
For this reason we here only consider systems of up to $N=256$ lattice
sites. It is worth noting that for a given $N$  
the RG method with the approximation as we use it is still much
faster than the ``numerically exact'' DMRG.
 
An alternative way to determine $E_0(\phi)$ would be to use Eq.\
(\ref{energyfromprop}) with the mean-field expectation values replaced
by the ones determined at $\Lambda=0$. Formally the use of the flow
equation for $\gamma_0$ and this method are
correct to leading order in $U$ but it turned out that calculating
$E_0(\phi)$ from $\gamma_0^{\Lambda=0}$ 
gives much better results compared to the accurate DMRG
data. For the results presented in the next section we thus used
Eqs.\ (\ref{gamma0flow}) and (\ref{energyfromOm}). There we will
demonstrate  that the RG data agree quantitatively with the DMRG data
for $U$ as large as the band width.

\section{Results}
\label{secresults}

\begin{figure}[hbt]
\begin{center}
\vspace{0.5cm}
\leavevmode
\epsfxsize8.2cm
\epsffile{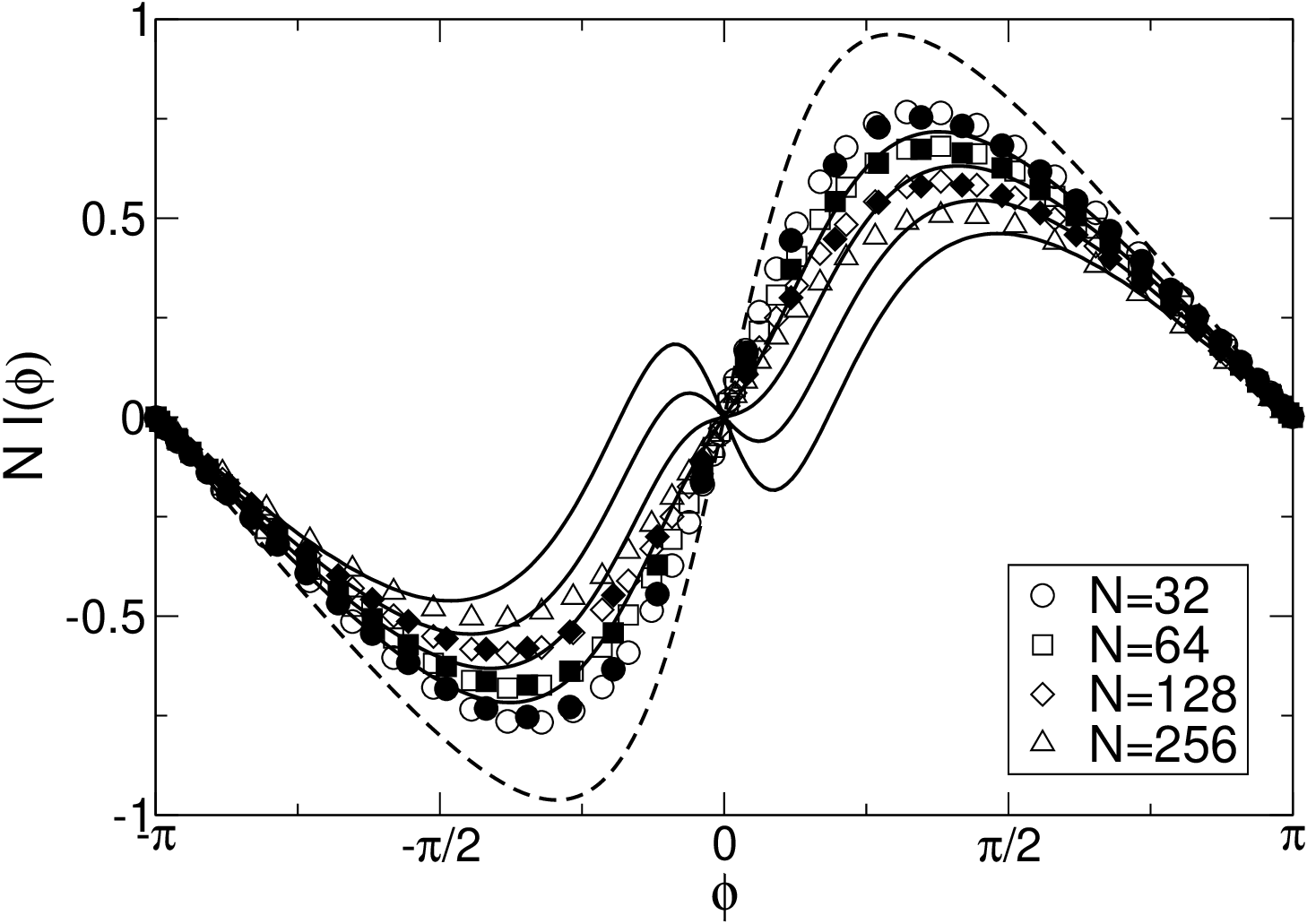}
\caption{Persistent current $N I(\phi)$ for $U=1$, $\rho=0.5$, and
  different $N$. The filled symbols are DMRG data, the
  open symbols RG data, and the solid lines are obtained from a
  Hartree-Fock calculation (for the same $N$). 
  The dashed line is the $U=0$ result Eq.\ (\ref{U0current}).}
\label{fig3}
\end{center}
\vspace{-0.0cm}
\end{figure}

\begin{figure}[hbt]
\begin{center}
\vspace{0.0cm}
\leavevmode
\epsfxsize8.2cm
\epsffile{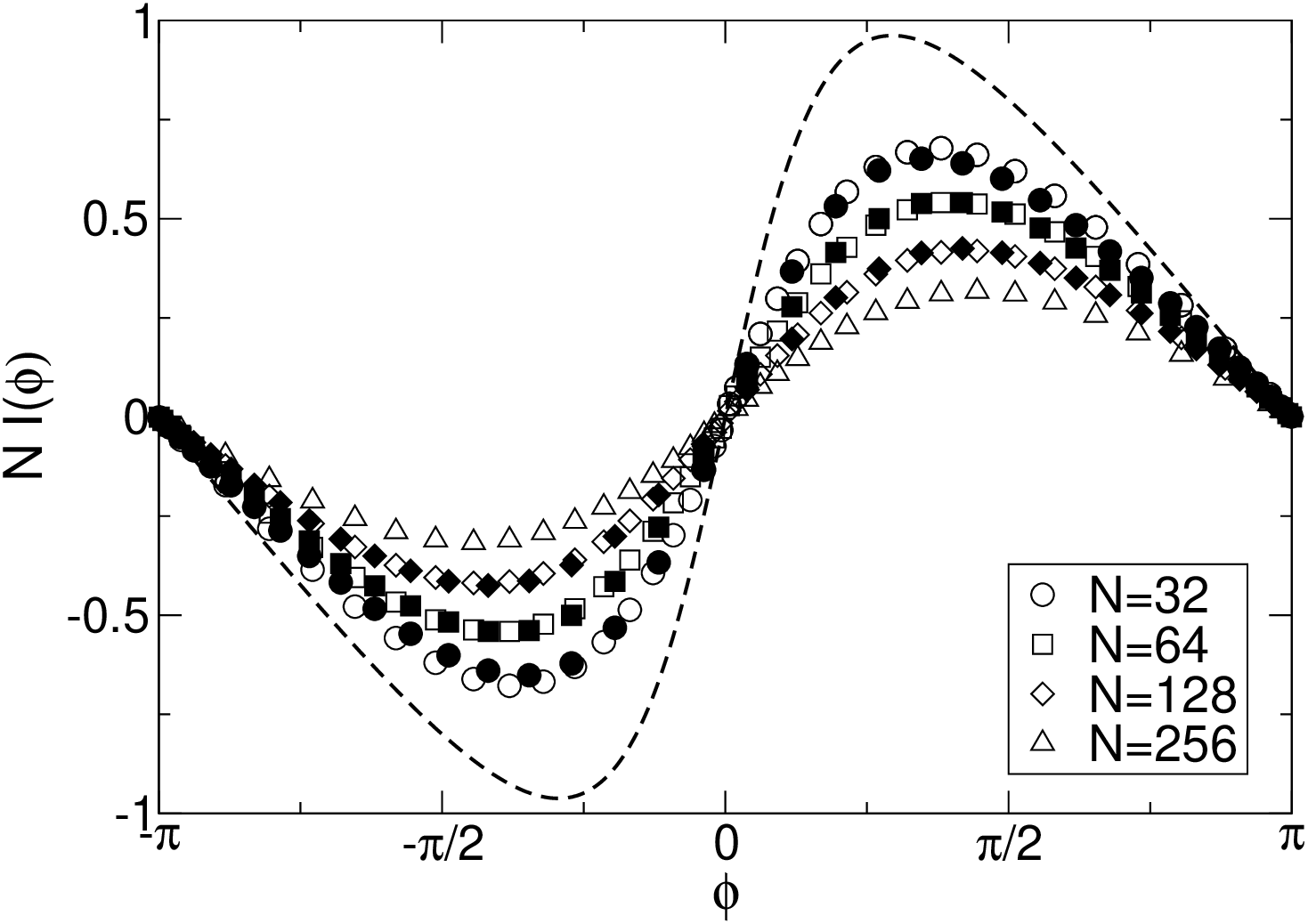}
\caption{The same as in Fig.\ \ref{fig3}, but for $U=1.73333$, $\rho=0.5$ and
without Hartree-Fock results.}
\label{fig4}
\end{center}
\vspace{-0.0cm}
\end{figure}

\begin{figure}[hbt]
\begin{center}
\vspace{0.0cm}
\leavevmode
\epsfxsize8.2cm
\epsffile{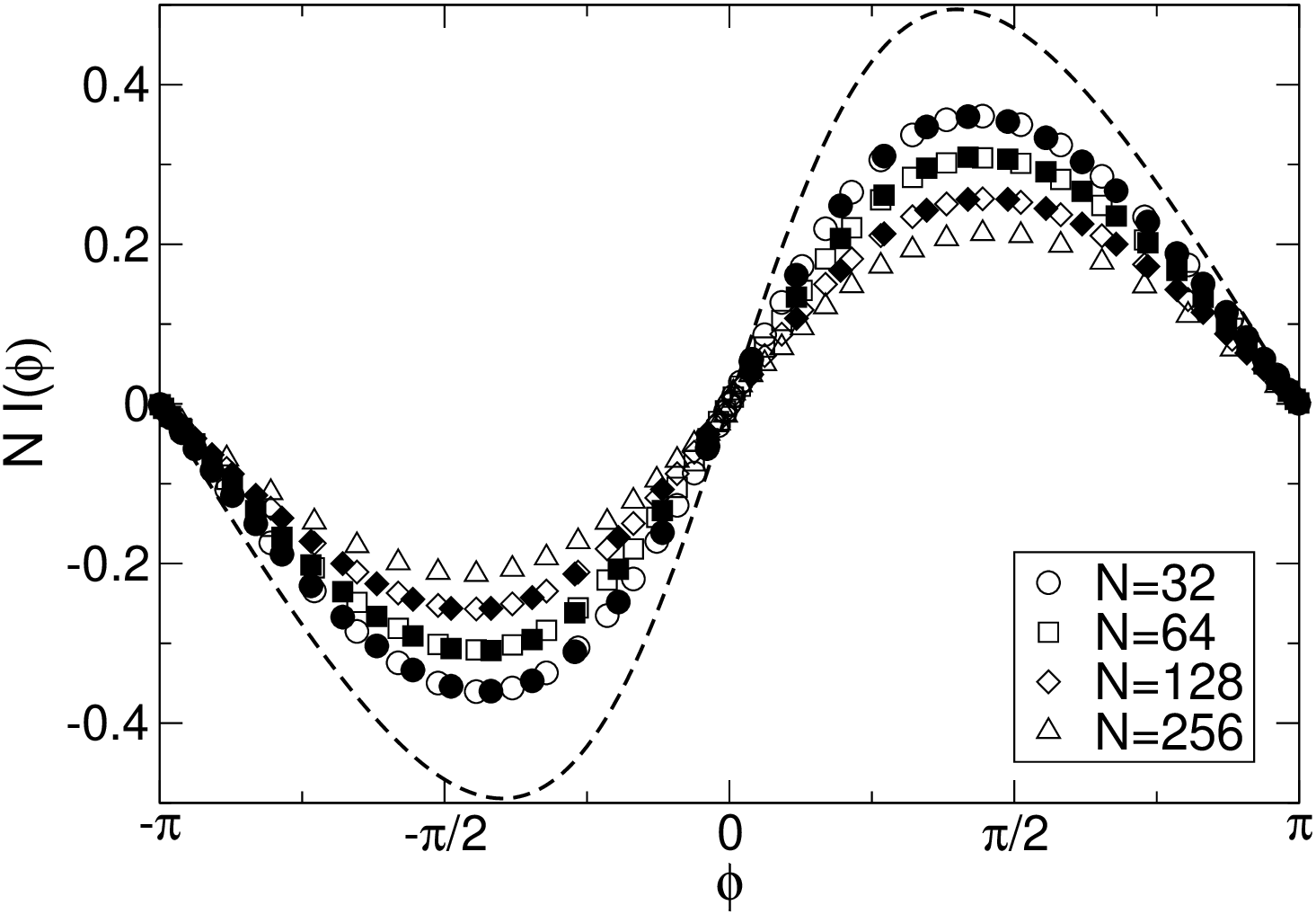}
\caption{The same as in Fig.\ \ref{fig3}, but for $U=1$, $\rho=0.25$ and
without Hartree-Fock results.}
\label{fig5}
\end{center}
\vspace{-0.0cm}
\end{figure}

In this section we present our results for the persistent current
including interaction and a hopping impurity obtained from DMRG and RG.
To demonstrate that the Hartree-Fock approximation should not be used in the present context 
we also show results obtained using this approximation. We here exclusively consider
$n=1/2$. In the DMRG and RG we have
calculated $E_0(\phi)$ for $0< \phi \leq \pi$ at $20$ different
$\phi_l$ and used the symmetry of the data to extend them to the 
periodicity interval $-\pi < \phi \leq \pi$. From these data we have
determined $I(\phi)$ using numerical differentiation. To make sure
that the error obtained from this is small we have not only used
simple centered differences but also applied an 
approximation scheme using Chebychev polynomials.\cite{numrep} 
This explains why the $\phi_l$ are
not equidistant (see Figs.\ \ref{fig3} to \ref{fig5}). Up to very small differences
both methods give the same current.

In Fig.\ \ref{fig3} $N I(\phi)$ is presented for $U=1$ and $\rho=0.5$,
which for $U=0$ corresponds to a transmission
$|T(k_F)|^2=0.64$, and different $N$. Additionally the leading
$1/N$ behavior at $U=0$ [Eq.\ (\ref{U0current})] is shown as a dashed line. 
As is obvious the interplay of interaction and impurity leads to a
current which vanishes faster than $1/N$. The DMRG and RG data agree 
quantitatively, which shows that the functional RG provides a very good
approximation (for a fairly large $U$) even if the flow of the vertex 
is neglected. The Hartree-Fock approximation gives qualitatively wrong
results. This shows that correlation effects not taken into account by
Hartree-Fock are very important. We thus do not consider this
approximation any further. For $U=1.733333$, i.e.\ close to the phase
transition into a charge density wave groundstate, the DMRG and RG data are shown 
in Fig.\ \ref{fig4} for the same $\rho$. For this larger $U$ the suppression is even
stronger but the RG data still show an excellent agreement with the high
precision DMRG results.

As discussed in the introduction in the large $N$ limit bosonization predicts 
that $I(\phi)$ is suppressed by a factor $N^{-\alpha_B-1}$ instead of $N^{-1}$
and proportional to $\sin{\phi}$. With increasing $U$, $\alpha_B$ gets larger and 
one thus expects this asymptotic regime to be reached faster.
Even though we observe a suppression
which is stronger than $N^{-1}$ from Figs.\ \ref{fig3} and \ref{fig4} 
it is not clear whether  a power-law behavior is realized. Furthermore the
data even for $U=1.73333$  are still far from being proportional to $\sin{\phi}$.
For a stronger impurity, i.e.\
smaller $\rho$ the asymptotic regime should also be reached faster
since already 
the non-interacting current 
is closer to a sinusoidal shape. This is demonstrated in Fig.\
\ref{fig5} for $U=1$ and $\rho=0.25$. The $U=0$ transmission for
$\rho=0.25$ is 
$|T(k_F)|^2=0.22$, i.e.\ this case corresponds to an already fairly
strong (bare) impurity.


\begin{figure}[hbt]
\begin{center}
\vspace{0.5cm}
\leavevmode
\epsfxsize7.7cm
\epsffile{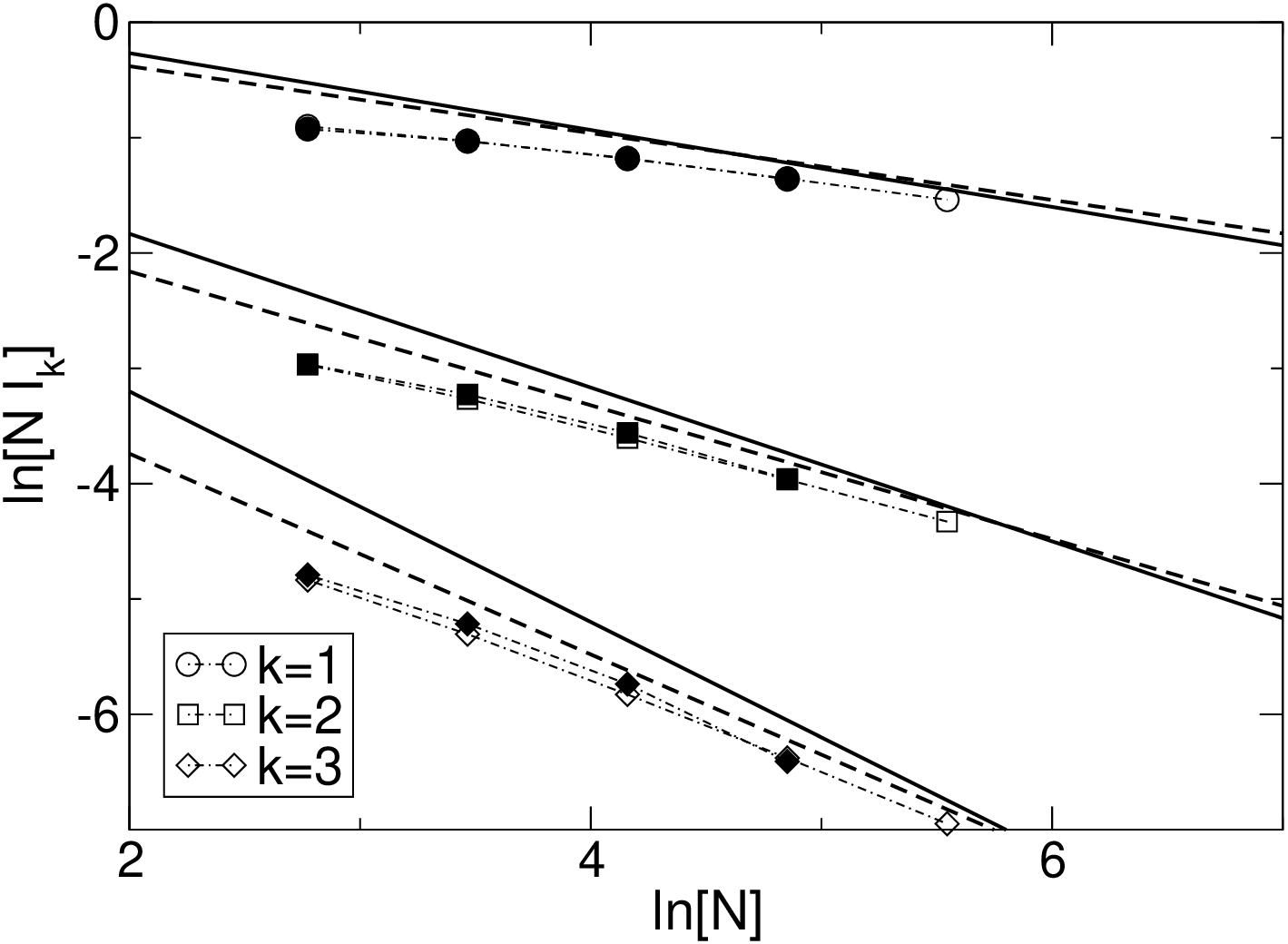}
\caption{$\ln{[N I_k]} $ as a function of $\ln{[N]}$ 
  for $U=1$, $\rho=0.25$, and $k=1,2,3$. 
  The filled symbols are DMRG data and the open symbols RG data. 
  The solid lines have slope $-k \alpha_B$ and the dashed ones 
  slope $-k \alpha_B^{\rm RG}$. The dashed-dotted lines are guides to
  the eyes. Open symbols which are not seen are hidden by the filled ones.} 
\label{fig6}
\end{center}
\vspace{-0.0cm}
\end{figure}

\begin{figure}[hbt]
\begin{center}
\vspace{0.0cm}
\leavevmode
\epsfxsize7.7cm
\epsffile{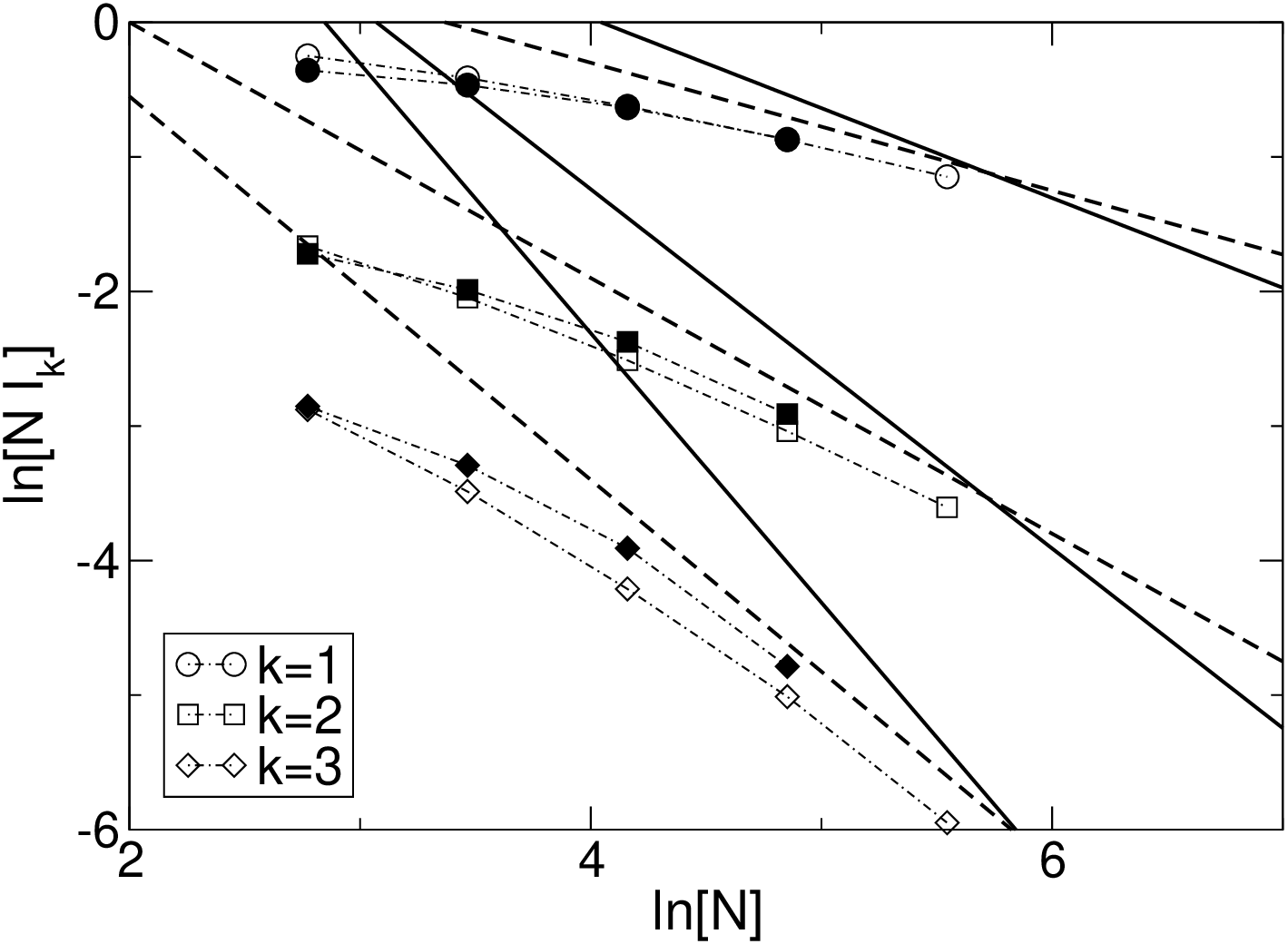}
\caption{The same as in Fig.\ \ref{fig6}, but for $U=1.73333$,
  $\rho=0.5$.}
\label{fig7}
\end{center}
\vspace{-0.0cm}
\end{figure}

\begin{figure}[hbt]
\begin{center}
\vspace{0.0cm}
\leavevmode
\epsfxsize7.7cm
\epsffile{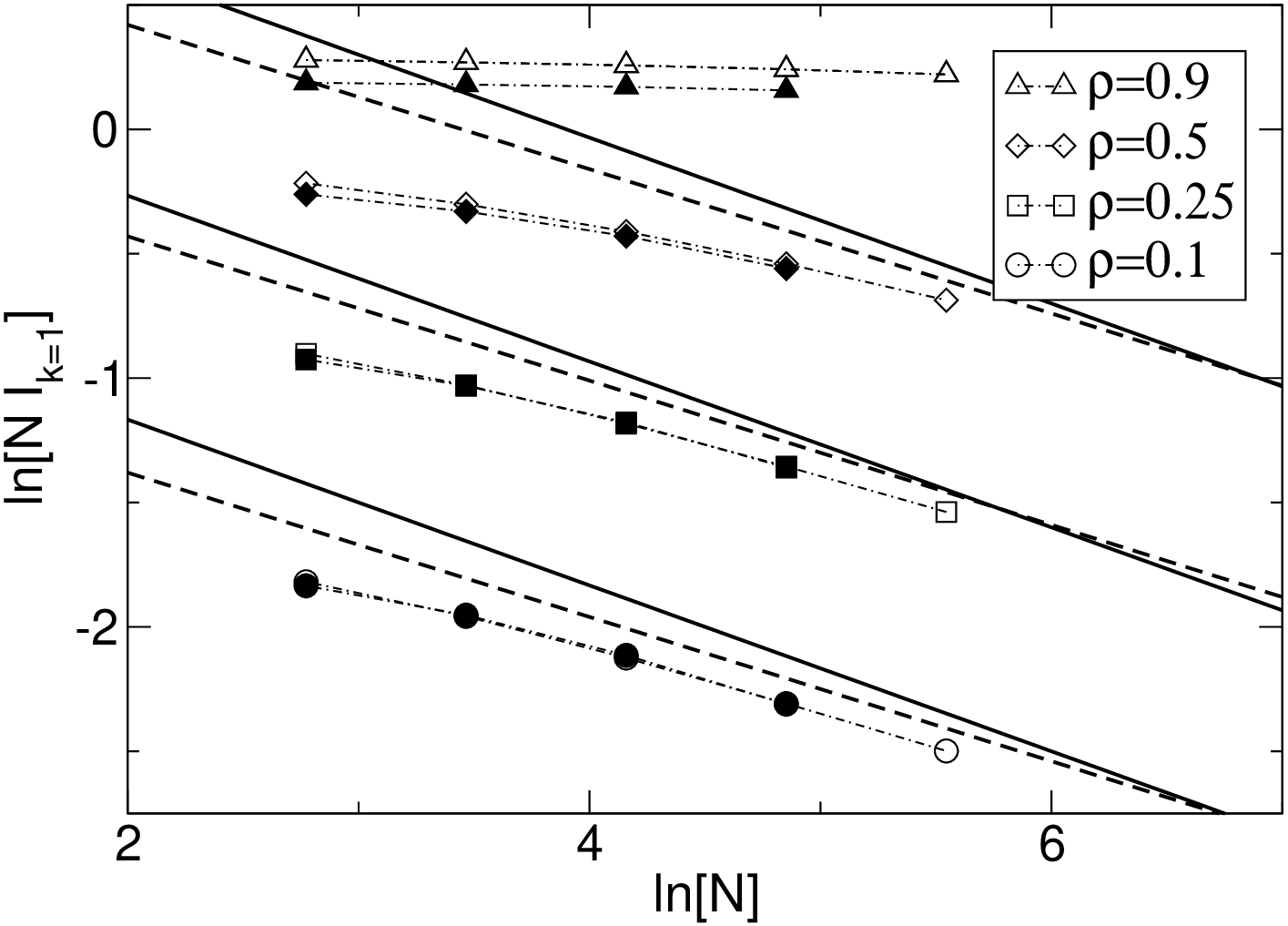}
\caption{The same as in Fig.\ \ref{fig6}, but for $U=1$, $k=1$, and
  different $\rho$.}
\label{fig8}
\end{center}
\vspace{-0.0cm}
\end{figure}

To analyze our data in more detail we have numerically 
determined the coefficients $I_k$ (for $k=1,2,3$) of a Fourier expansion 
\begin{eqnarray}
I(\phi) = \sum_{k=1}^{\infty}  I_k \sin{(k\phi)}
\label{fourierexp}
\end{eqnarray}
for different $U$, $\rho$, and $N$. 
From bosonization one expects, that $N I_k$ for $k=1$ decays 
asymptotically as $N^{-\alpha_B}$ and that the higher order Fourier
coefficients die off even faster. To better understand the behavior 
of $N I_k$ for $k=2$ and $3$ we will use the one-particle picture
mentioned in the introduction. According to this many bosonization results can be
understood in terms of an effective $|T(k_F)|$ which is suppressed by a 
factor $N^{-\alpha_B}$ compared to the non-interacting transmission amplitude. 
Using this and Eq.\ (\ref{U0currentexp}) one concludes that $N I_k$ (at least for $k=2$
and 3) should 
decay as $N^{-k \alpha_B}$. This argument
is based on the use of
a single particle picture and needs further justification. 
Fig.\ \ref{fig6} shows $\ln{[N I_k]}$ ($k=1,2,3$) as a function of $\ln{[N]}$
for $U=1$, $\rho=0.25$ and Fig.\ \ref{fig7} contains
data for $U=1.73333$, $\rho=0.5$. 
In addition to the DMRG and RG data straight lines with slope $- k
\alpha_B$ [see Eq.\ (\ref{Krhohalf})] (solid lines) and $- k
\alpha_B^{\rm RG}$ (dashed lines) for $k=1,2,3$ are shown. Since the
flow of the two-particle vertex has been neglected in the functional
RG, within this approach we can only expect exponents to be correct to leading
order in $U$. $\alpha_B^{\rm RG}$ is the boundary exponent of the
local spectral weight (close to the boundary) as we have determined it
in Refs.\ \onlinecite{VMa} and \onlinecite{VMb} using the RG for much
larger system sizes. As expected it agrees with $\alpha_B$ only to
leading order in $U$. For $U=1$ we have  $\alpha_B^{\rm RG} =0.291$
instead of $\alpha_B=1/3$ and for $U=1.73333$,   $\alpha_B^{\rm RG}
=0.475$ instead of $\alpha_B=0.667$.
From both figures it is clear that higher order Fourier coefficient
decay faster than lower ones. On the other hand only the data obtained
for a already fairly strong bare impurity $\rho=0.25$ follow at large
$N$ lines with the expected exponents (see Fig.\ \ref{fig6}). Even in
this case the DMRG data for larger $N$ have to
bend further down to
reach the predicted behavior (slope $\sim - k \alpha_B$). 
For large $N$ also the behavior of the $k=2$ and $3$ Fourier coefficients seems to
be consistent with the above predictions obtained using the
one-particle language. 
For $\rho=0.5$ - an impurity of
intermediate strength - even for large $U=1.73333$ and up to $N=128$
respectively $N=256$
lattice sites the DMRG and RG data still show a strong curvature. For
this case the bosonization prediction cannot be demonstrated
unambiguously although the data show a tendency towards a behavior
which is consistent with this prediction. 
Results for $U=1$, $k=1$,
and different $\rho$ are summarized in Fig.\ \ref{fig8}. The bare
transmission for $\rho=0.9$ is $|T(k_F)|^2=0.99$ at $U=0$ and for
$\rho=0.1$ we have $|T(k_F)|^2=0.04$. These findings are consistent
with the results obtained for a local
observable - the spectral weight close to the impurity.\cite{VMa,VMb}


One can stress the single particle picture even further and use the
non-interacting expression for $I(\phi)$ [Eq.\ (\ref{U0current})] with an
effective transmission amplitude $|T_{\rm eff}(k_F)|$ as a parameter 
to fit our data for different $U$, $\rho$, and $N$. For small $N$ this
fitting procedure cannot be expected to work as good as for larger
ones since Eq.\ (\ref{U0current}) only provides the large $N$ behavior
of the current in the non-interacting case. One then expects that $|T_{\rm
  eff}(k_F)|$ for $N \to \infty$ scales as $N^{-\alpha_B}$ (respectively
$N^{-\alpha_B^{\rm RG}}$ for the RG data). Similar to Fig.\ \ref{fig8}
the effective transmission amplitudes show this behavior  only for 
small $\rho$ and large $N$. For intermediate to weak impurities the
scaling limit is not reached. We have demonstrated that
the use of the single particle language can be useful in the present
context. Nevertheless it involves arguments which are ``hand waving''
and we thus do not follow this route any further.

\section{Summary and perspectives}
\label{secsummary}

With our work we achieved two goals. We first 
determined the detailed functional form of the persistent current
$I(\phi)$ realized in a lattice model of one-dimensional
interacting electrons penetrated by a magnetic flux in the presence of
a single impurity. This enabled us to go beyond the large $N$ scaling 
limit in which the current is proportional to  $\sin{\phi}$ and vanishes
as $N^{-\alpha_B-1}$, and investigate the corrections to this
behavior. 
Only for large bare impurities and large system sizes we were
able to confirm the bosonization prediction. For intermediate to weak 
impurities the
asymptotic limit described by bosonization is not reached for systems
of up to $256$ lattice sites and the
current shows a more complex behavior. These results 
are consistent with our observations for the local spectral
weight close to an impurity.\cite{VMa,VMb} 
By comparing the high precision DMRG data to the ones obtained within a
functional RG approach, where the flow of the two-particle vertex was
neglected, we moreover demonstrated that the latter method can to a very high
accuracy  also be used for global observables such as the persistent current.
This holds for interactions as large as the band width. 
Our results show that even in the cases where the universal scaling limit is not
reached correlation effects included in the RG are still very
important. This is demonstrated by a comparison to results obtained
with the Hartree-Fock approximation which are qualitatively wrong.

Besides the hopping impurities we have 
also studied the case of a single site impurity\cite{VMa} for selected parameter sets
and obtained results which are comparable to the ones presented above. 

The functional RG can be applied to other microscopic
models (including e.g.\ long range interaction), at finite temperatures, and for
models with disorder. We expect that
including the flow of the two-particle vertex 
- as discussed for the local spectral weight in Ref.\ \onlinecite{Sabine} -
will further increase the accuracy of the RG data for the
persistent current. The flow of the interaction vertex is of special 
importance for models with spin 
since in Luttinger liquids with spin the flow of the electron-electron 
backscattering has to be taken into account.\cite{Solyom} It should thus be 
included in further applications of the RG method on one-dimensional 
models. Considering currents driven by a small difference in
voltage within the RG it is also possible to include leads and to
realize a great variety of electronic circuits. 
It can be expected that in small conducting molecular systems 
of the size of a few ten to a few hundred lattice sites
which in the near future will be accessible to experiments
correlation effects will be important but the asymptotic limit
described by the effective field theory will not be reached. 
The RG is an ideal tool to investigate the electronic properties of such systems.

\section*{Acknowledgements}
We would like to thank M.\ Henkel, H.\ Schoeller, W.\ Metzner, and K.\ Sch\"onhammer 
for very valuable discussions and P.\ Schmitteckert for comments on
the manuscript.  U.S.\ is grateful to the Deutsche Forschungsgemeinschaft 
for support from the Gerhard-Hess-Preis and V.M.\ acknowledges support from the 
Bundesministerium f\"ur Bildung und Forschung (Juniorprofessor Programm).


\begin{thebibliography}{*}
\bibitem{Schick} M.\ Schick, Phy.\ Rev.\ {\bf 166}, 404 (1968). The
  author only considers the case of the one-dimensional electron gas
  with quadratic dispersion. This model is Galilean-invariant 
  which implies that $v_J = v_F$ independent of the interaction, 
  where $v_F$ is the Fermi velocity.
\bibitem{footnote3} There is a misprint in Eq.\ (3) of Ref.\
  \onlinecite{VMb}. The first minus sign on the right hand side has to
  be replaced by a plus.
\end{thebibliography}
\end{document}